\documentclass{mva_style}
\usepackage{graphicx}

\usepackage{algorithm}
\usepackage{algorithmic}
\usepackage{latexsym}
\usepackage{cite}
\usepackage{enumitem}
\usepackage{amssymb,mathrsfs}
\usepackage{booktabs}
\usepackage{microtype}
\usepackage{siunitx}
\usepackage{dcolumn}
\usepackage{colortbl}
\usepackage{subfigure}
\usepackage{float}
\usepackage{color}
\usepackage[dvipsnames]{xcolor}
\usepackage{amsmath}
\usepackage{booktabs}
\usepackage{multirow}

\finalcopy 

\begin{document}
\title{A survey of manifold learning and its applications for multimedia}

\author{
  Hannes Fassold\\
  JOANNEUM RESEARCH - DIGITAL\\
  {\tt hannes.fassold@joanneum.at}\\
}

\maketitle

\section*{\centering Abstract}
\textit{
  Manifold learning is an emerging research domain of machine learning. In this work, we give an introduction into manifold learning and how it is employed for important application fields in multimedia.
}


\section{Introduction}
\label{sec:manisurvey:intro}

Deep learning methods are nowadays the best way for the automatic analysis of multimedia data (e.g. images, video or 3D data) for tasks like classification or detection. However, classic neural networks are restricted to data lying in vector spaces, while data residing in smooth non-Euclidean spaces arise naturally in many problem domains. For example, a 360$^{\circ}$ camera actually captures a spherical image, not a rectangular image. We will focus in the following on manifolds, especially Riemannian manifolds, which are well suited for generalizing a vector space because they are locally Euclidian and differentiable. 

\begin{figure}[t]
    \centering
    \includegraphics[width = 0.33\textwidth]{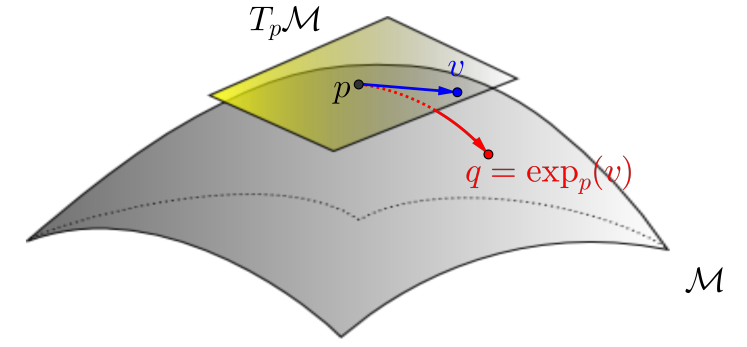}
    \caption{Tangent space and exponential map on a 2-dimensional manifold. Image courtesy of \cite{Miranda2017JR}.}
    \label{fig:jr:tangent_plane}
    \vspace{-5mm} 

\end{figure}

A \emph{manifold} $M$ of dimension $d$ corresponds to a topological structure which locally (so in the neighborhood of a point $\boldsymbol{p}\in M$) looks like a $d-$dimensional Euclidean space.
The "best" local approximation of this neighborhood of $\boldsymbol{p}$ with a $d-$dimensional Euclidean space is its \emph{tangent space} $T_p M$. The tangent space $T_p M$ can be seen as a linear approximation of $M$ around $\boldsymbol{p}$. For example, for a 2-dimensional manifold its tangent space $T_p M$ is the tangent plane going through this point (see Figure \ref{fig:jr:tangent_plane}). A \emph{Riemannian manifold} is a smooth manifold $M$ equipped with a positive definite inner product $g_p$ on the tangent space $T_p M$ of each point $\boldsymbol{p}$. 

The inner product $g$ induces a norm on the tangent space, which subsequently allows us to calculate curve lengths and distances on the manifold $M$. For each curve $c(t)$ on the manifold its length can be calculated by integrating the norm along the curve (for details see \cite{Fletcher2010JR, Porikli2007JR, Sommer2020JR, Hauberg2012JR, Calinon2020JR}). A \emph{geodesic} curve is a \emph{length-minimizing} curve connecting two  points $\boldsymbol{p}$ and $\boldsymbol{q}$ on the manifold. The distance between these points is defined as the length of the geodesic.

Let $\boldsymbol{p}$ be a (reference) point on the manifold and $v$ a vector of its tangent space $T_p M$. The vector $v$ can be mapped now to the point $\boldsymbol{q}$ on the manifold that is reached after unit time $t=1$ by the geodesic $c(t)$ starting at $\boldsymbol{p}$ with tangent vector $v$. This mapping $exp_p(v): T_p M \rightarrow M$ is called the exponential map at point $\boldsymbol{p}$. 

The inverse mapping $log_p(\boldsymbol{q}): M \rightarrow T_p M$ is uniquely defined around a neighborhood of $\boldsymbol{p}$. Informally, the exponential map and logarithm map move points back and forth between the manifold and the tangent space (see Figure \ref{fig:jr:tangent_plane}) while preserving distances. Furthermore, derivative operators like \emph{differential}, \emph{intrinsic gradient}, \emph{divergence} and \emph{laplacian} can be also defined on a manifold \cite{Bronstein2017JR, Tu2007JR}, which allows us to perform calculus on the manifold.

Closely related to manifolds are Lie groups. A \emph{Lie group} is a smooth manifold that also forms a \emph{group} \cite{Fletcher2010JR}, where both group operations (commonly called \emph{multiplication} and \emph{inverse}) are smooth mappings of manifolds. The \emph{Lie algebra} $\mathfrak{g}$ of a Lie group $M$ is defined as the tangent space at the identity  $T_e M$, where $e$ is the identity element of the group (see section $16$ in \cite{Tu2007JR}).

Key components of neural networks -- like mean, convolution, nonlinearities and batch normalization --  can be defined on Riemannian manifolds as described in \cite{Chakraborty2022JR, Zhen2019JR, Bouza2020JR, Lou2020JR, Chakraborty2020JR}. Optimization algorithms for Riemannian manifolds (gradient descent, SGD, Adam etc.) can be found in \cite{Sim2021JR, Cho2017JR, Lezcano2019JR, Alimisis2021JR, Li2020JR, Becigneul2019JR, Absil2007JR, Kasai2019JR}.

Commonly encountered examples of Riemannian manifolds in computer vision are the $n-$sphere $S^n$, the manifold of $n \times n$ symmetric positive matrices $P_n$, the special orthogonal group $SO(n)$ (rotation matrices), the special euclidean group $SE(n)$ (rigid body transformations), Grassman manifold $Gr(n, p)$ (collection of all $p-$dimensional linear subspaces in $\mathbb{R}^n$, see \cite{Bendokat2020JR}) and the Stiefel manifold $St(n, p)$ (collection of all $p$-dimensional orthogonal bases in $\mathbb{R}^n$).

In the following, we will give an overview of manifold learning methods employed in important application fields in multimedia (similarity search, image classification, synthesis \& enhancement, video analysis, 3D data processing, nonlinear dimension reduction) and about available open source software frameworks.


\section{Similarity search \& retrieval}
\label{sec:manisurvey:retrieve}
Image retrieval deals with searching for similar images in an image gallery, given a certain query image (see the surveys \cite{Dubey2021JR, Chen2022JR}). Many methods employ for this \emph{metric learning}, which transforms input images into \emph{embeddings} ($\approx$ feature vectors) and learns a distance function between these embeddings.

The authors of \cite{Bai2017JR} propose \emph{regularized ensemble diffusion} for refining/reranking the initial similarity search results. They show that regularized ensemble diffusion is significantly more robust against noise in the data than standard diffusion. A \emph{diffusion} process \cite{Donoser2013JR} models the relationship between objects on a graph-based manifold, wherein similarity values are diffused along the geodesic path in an iterative way.

In \cite{Iscen2018JR} an unsupervised framework is presented for the identification of \emph{hard training examples} for the training of an embedding. Hard training examples (both positive and negative samples) are identified by disagreement between euclidean and manifold similarities.

A time- and memory-efficient algorithm for estimating similarities on the data manifold is proposed in \cite{Aziere2019JR}. They adapt the random walk procedure to estimate manifold similarities only an a small number of data in each mini-batch, rather than on all training data.

The $MLS^{3}RDUH$ algorithm \cite{Tu2020JR} utilizes the intrinsic manifold structure in the feature space and cosine similarity to reconstruct the local semantic structure and build a similarity matrix upon it. Then a novel \emph{log\hbox{-}cosh} hashing loss function is used to optimize the hashing network in order to generate compact hash codes, guided by the similarity matrix.

The work of \cite{Dutta2020JR} proposes a unsupervised metric learning algorithm that learns a metric in a lower dimensional latent space using constraints provided as tuples, which rely on pseudo-labels obtained by a graph-based clustering method (\emph{authority ascent shift}). The parameters of the approach are learned jointly using Riemannian optimization on a product manifold.


\section{Image classification \& object detection}
\label{sec:manisurvey:class}

The work \cite{Lezcano2019JR} proposes a framework for the transformation of problems with manifold constraints into unconstrained problems on an Euclidean space through a mechanism they call \emph{dynamic trivializations}. They show how to implement these trivializations efficiently for a large variety of commonly used matrix manifolds and provide a formula for the gradient of the matrix exponential.

The authors of \cite{Verma2019JR} propose \emph{manifold mixup}, a novel regularizer which forces the training to interpolate between hidden representations -- captured in the intermediate layers of the network -- of samples. It can be seen as a generalization of input mixup which does the interpolation on a random layer of the network (whereas input mixup uses always layer 0). Experiments for the task of image classification show that manifold mixup flattens the class-specific representation (lower variance) and generates a smoother decision boundary.

In \cite{Atigh2021JR} \emph{Hyperbolic Busemann learning} with ideal prototypes is introduced. It places class prototypes at the ideal boundary of the \emph{Poincare ball} (a hypersphere manifold with hyperbolic geometric) and introduces the \emph{penalized Busemann loss} for optimizing with respect to ideal prototypes. They prove its equivalence to logistic regression for the one-dimensional case.

An approach for few-shot image classification is presented in \cite{Rodriguez2020JR} which proposes \emph{embedding propagation} as an unsupervised non-parametric regularizer. Embedding propagation leverages interpolation between the extracted features of a neural network, based on a similarity graph. Experiments show that embedding propagation yields a smoother embedding manifold and gives better performance on three standard datasets for few-shot image classification.

\begin{figure}[t]
    \centering
    \includegraphics[width = 0.38\textwidth]{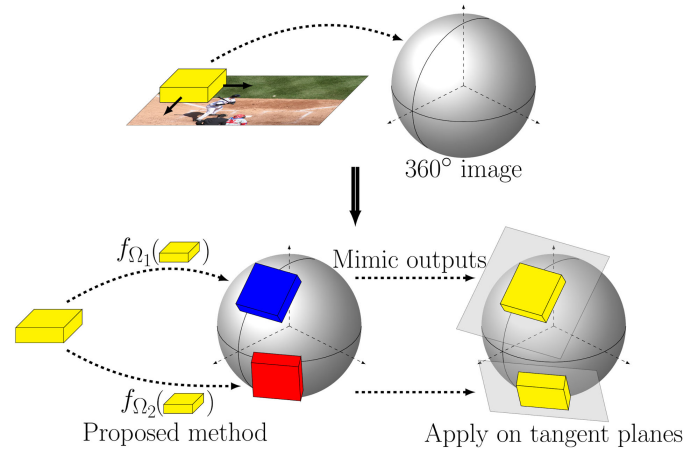}
    \caption{Transfer CNNs trained on flat images to 360$^{\circ}$ images with the method from \cite{Su2022JR}.}
    \label{fig:jr:spherical_convolution}
    \vspace{-5mm} 
\end{figure}

The work \cite{Su2022JR} introduces a knowledge distillation method which is able to transfer an existing CNN model trained on perspective images to \emph{spherical} images captured with a 360$^{\circ}$ camera \emph{without} any additional annotation effort (see Figure \ref{fig:jr:spherical_convolution}). They train a spherical Faster R-CNN model with this method, demonstrating that a object detector for spherical images (in equirectangular projection) can be trained without any annotations in the 360$^{\circ}$ images.

\section{Image synthesis \& enhancement}
\label{sec:manisurvey:enhance}

For image synthesis and enhancement, state of the art algorithms employ either GANs (\emph{generative adversial networks} \cite{Wang2021JR}) or \emph{diffusion models} \cite{Croitoru2022JR}.

The authors of \cite{Chung2022JR} show that current solvers employed in diffusion models throw the generative sample path off the data manifold, causing the error to accumulate. They propose an additional correction term inspired by the manifold constraint to force the iterations to be close to the data manifold. The proposed manifold constraint is easy to add to a solver, yet boosts its performance significantly.

In \cite{Dai2022JR} a novel implicit data augmentation approach for training GANs is proposed which facilitates stable strong and synthesizes high-quality samples. Specifically, the discriminator is interpreted as a metric embedding of the real data manifold, which offers real distances between real data samples. Experiments show that the proposed method improves the performance of image synthesis in the low-data regime.

A method for comparing data manifolds based on their topology is presented in \cite{Barannikov2021JR}. They introduce novel tools, specifically \emph{cross-barcode} and \emph{manifold topology divergence score}, which are able to track spatial discrepancies between manifolds on multiple scales. They apply it to assess the performance of generative models in various domains (images, 3D shapes or time series) and demonstrate that these tools are able to detect common problems of GAN-based image synthesis like mode dropping, mode collapse and image disturbance.

\begin{figure}[t]
    \centering
    \includegraphics[width = 0.42\textwidth]{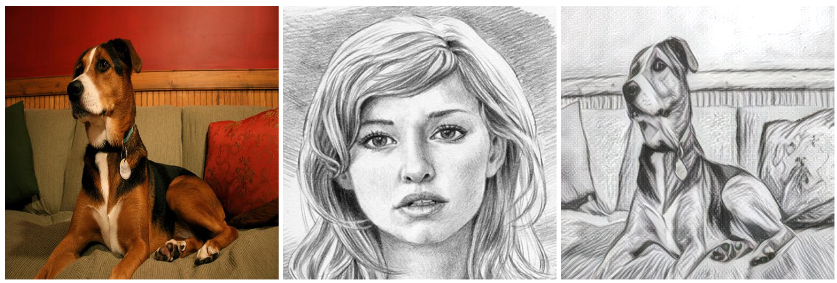}
    \caption{From left to right: Content image, style image, style-transferred image \cite{Luo2022JR}.}
    \label{fig:jr:style_transfer_result_pama}
    \vspace{-5mm} 
\end{figure}

The work \cite{Luo2022JR} proposes \emph{progressive attentional manifold alignment} for style transfer, which progressively aligns content manifolds to their most related style manifolds. Afterwards, \emph{space-aware interpolation} is performed in order to increase the structural similarity of the corresponding manifolds, which makes it easier for the attention module to match features between them. Experiments show that the method generates high-quality style-transferred images (see Figure \ref{fig:jr:style_transfer_result_pama}).

The authors of \cite{Ramasinghe2021JR} proposes an algorithm for improving the diversity and visual quality of images generated by a conditional GAN, by systematically encouraging a \emph{bi\nobreakdash-lipschitz} mapping between the latent and output manifold. The performance improvement is shown on several image\nobreakdash-to\nobreakdash-image translation tasks, like landmark\nobreakdash-to\nobreakdash-face or sketch\nobreakdash-to\nobreakdash-anime.

\begin{figure}[t]
    \centering
    \includegraphics[width = 0.47\textwidth]{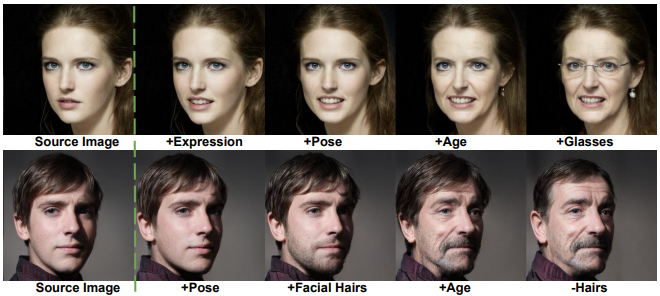}
    \caption{Image editing with FLAME \cite{Rishubh2022JR}.}
    \label{fig:jr:flame_algorithm}
    \vspace{-3mm} 
\end{figure}

The FLAME algorithm proposed in \cite{Rishubh2022JR} performs highly realistic image manipulations (e.g. changing expression, hair style or age of a synthetic face, see Figure \ref{fig:jr:flame_algorithm}) with minimal supervision. It estimates linear latent directions in the latent space of \emph{StyleGAN2} using only a few image pairs and introduces a novel method for sampling from the attribute style manifold.

\section{Video analysis}
\label{sec:manisurvey:video}

Most manifold learning methods for video analysis deal with the important task of human action recognition. Often they employ neural networks over the manifold $P_n$ of symmetric positive matrices  (usually covariance matrices) for this.

The authors of \cite{Zhen2019JR} propose a \emph{dilated convolution} operator on manifolds, based on the \emph{weighted Frechet mean} \cite{Chakraborty2022JR}, as well as a \emph{residual connection} operator. Both are important building blocks of modern neural networks. They construct a manifold-valued network employing covariance matrices (calculated from CNN features) and train this network for human action detection on the UCF-11 video dataset.

In \cite{Bouza2020JR} the convolution is defined as the weighted sum (reprojected to the manifold) in the tangent space $T_a M$, where $a$ is the Frechet mean of the input points for the convolution. They show that their proposed convolution operator is an isometry of the manifold, which corresponds to the translation-invariance property of the convolution in an Euclidean space.

The work \cite{Friji2020JR} proposes a geometry-aware deep learning algorithm for skeleton-based action recognition, where skeleton sequences are modeled as trajectories on \emph{Kendall's shape space} and then fed into a CNN-LSTM network. Kendall's shape space \cite{Kendall1999JR, Guigui2021JR} is a special quotient  manifold that defines shape as the geometric information that remains when location, scaling and rotational effects are filtered out.

The algorithm \cite{Wang2022aJR} adopts a neural network over the manifold $P_n$ of symmetric positive definite matrices as the backbone and appends a cascade of \emph{Riemannian autoencoders} to it in order to enrich the information flow within the network. Experiments on the tasks of emotion recognition, hand action recognition and human action recognition demonstrate a favourable performance compared to state of the art methods.


\section{3D data processing}
\label{sec:manisurvey:3d}

\begin{figure}[t]
    \centering
    \includegraphics[width = 0.45\textwidth]{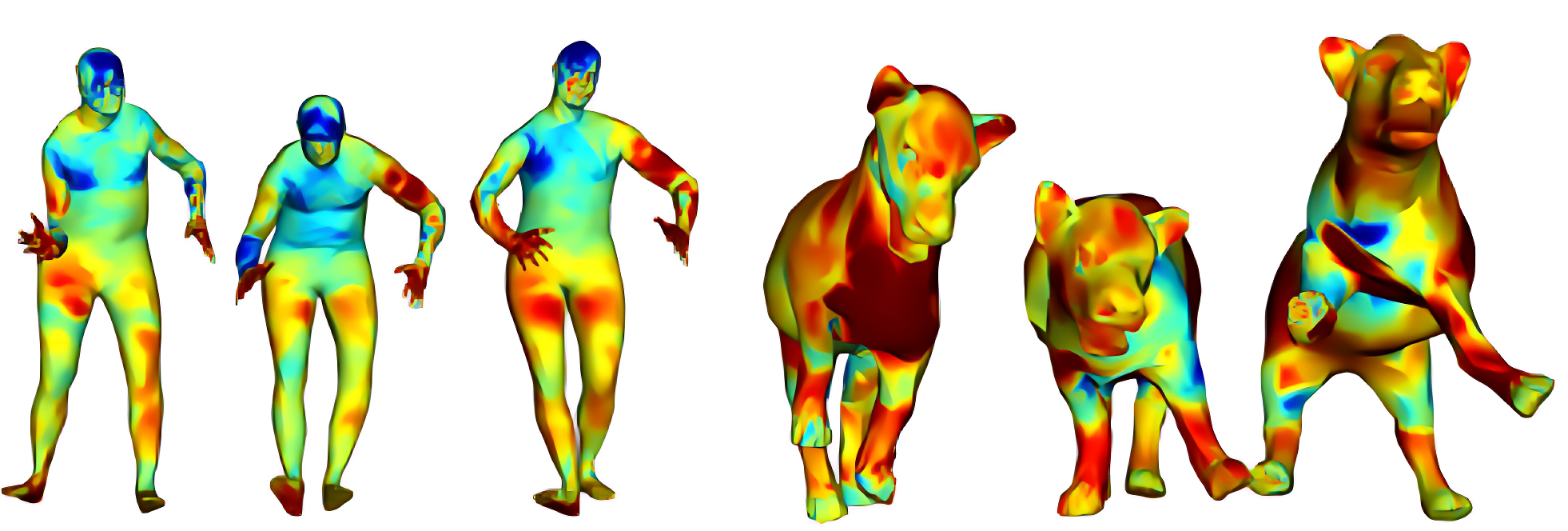}
    \caption{Generated 3D models with the geometric disentanglement algorithm from \cite{Tatro2022JR}.}
    \label{fig:jr:3d_models}
    \vspace{-5mm} 
\end{figure}

The work \cite{Tatro2022JR} proposes a novel algorithm for geometric disentanglement (separate intrinsic and extrinsic geometry) of 3D models, based on the fundamental theorem for surfaces. They describe surface features via a combination of \emph{conformal factors} and surface normal vectors and propose a convolutional mesh autoencoder based on these features. The conformal factor defines a conformal (angle-preserving) deformation between two manifolds. The algorithm achieves state-of-the-art performance on 3D surface generation, reconstruction and interpolation tasks (see Figure \ref{fig:jr:3d_models}).

The authors of \cite{Hamu2022JR} propose an approach for learning generative models on manifolds by minimizing the \emph{probability path divergence}. Unlike other continuous flow approaches, it does not require solving an ordinary differential equation during training. 

In \cite{Chen202aJR} a method for rotation (pose) estimation of 3D objects from point clouds and images is presented.  For this, they propose a novel \emph{manifold-aware} gradient in the backward pass of rotation regression that directly updates the neural network weights.

The work \cite{Koestler2022JR} introduces \emph{intrinsic neural fields}, a novel and versatile representation for neural fields on manifolds. Intrinsic neural fields are based on the eigenfunctions of the \emph{Laplace-Beltrami} operator, which can represent detailed surface information directly on the manifold. Furthermore, they extend \emph{neural tangent kernel analysis} to manifolds for better insight into the spectral properties of neural fields.


\section{Nonlinear dimension reduction}
\label{sec:manisurvey:ndr}

Many real world high-dimensional datasets are actually lying in a low-dimensional manifold (\emph{manifold hypothesis}). Nonlinear dimensional reduction algorithms project high-dimensional data onto such a low-dimensional manifold, while trying to preserve distance relationships in the original high-dimensional space as good as possible. 

Classical approaches for nonlinear dimension reduction are Isomap, Local Linear Embedding (LLE) and Laplacian Eigenmaps (see the survey in \cite{Cayton2005JR}). In recent years, more powerful approaches like \emph{t-SNE}, \emph{UMAP}, \emph{TriMAP} and \emph{PaCMAP} have emerged \cite{Wang2022JR}. From these, PaCMAP seems to preserve best both the global and local structure of the high-dimensional data. 

In \cite{Sarfraz2022JR}, the \emph{h-NNE} algorithm is proposed, which is competitive with t-SNE and UMAP in quality while being on order of magnitude faster. The significant runtime advantage is possible as h-NNE avoids solving an optimization problem and relies on \emph{nearest neighbor graphs} instead.

\begin{figure}[t]
    \centering
    \includegraphics[width = 0.45\textwidth]{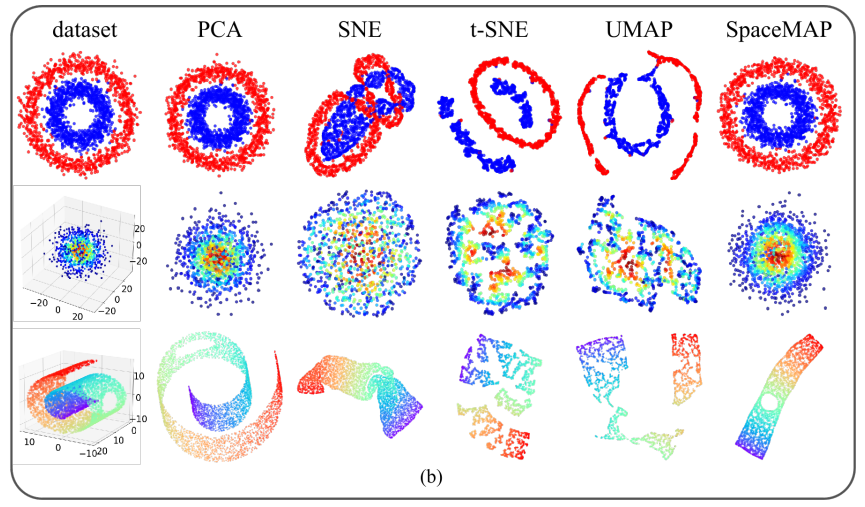}
    \caption{Comparison of classic nonlinear dimension reduction methods with SpaceMAP\cite{Zu2022JR}.}
    \label{fig:jr:spacemap}
    \vspace{-5mm} 
\end{figure}

The \emph{SpaceMAP} algorithm \cite{Zu2022JR} (see Figure \ref{fig:jr:spacemap}) introduces the concept of \emph{equivalent extended distance}, which makes it possible to match the capacity between two spaces of different dimensionality. Furthermore, \emph{hierarchical manifold approximation} is performed based on the observation that real-world data has often a hierarchical structure.

The \emph{DIPOLE} algorithm proposed in \cite{Wagner2022JR} corrects an initial embedding (e.g. calculated via Isomap) by minimizing a loss functional with both a local, metric term and a global, topological term based on \emph{persistent homology}. Unlike more ad hoc methods for measuring the shape of data at multiple scales, persistent homology is rooted in algebraic topology and enjoys strong theoretical foundations. 

For measuring the intrinsic dimension of a data distribution, in \cite{Tempczyk2022JR} a method is presented based on recent progress in likelihood estimation in high dimensions via \emph{normalizing flows}.

\section{Open source software frameworks}
\label{sec:manisurvey:software}
The Python packages \emph{Geomstats} \cite{Miolane2022JR, Miolane2020JR}, \emph{geoopt} \cite{Becigneul2019JR} and \emph{Pymanopt} \cite{Townsend2016JR} provide implementation of the standard operators (norm, distance, exp, log, retraction, parallel transport etc.) for commonly used  manifolds like $S^n$, $P_n$, $SO(n)$, $SE(n)$, $Gr(n, p)$ and $St(n, p)$. 

\emph{Geomstats} and \emph{geoopt}  support also more exotic manifolds like Birkhoff polytope \cite{Douik2018JR}, stereographic projection model, Kendall's shape space \cite{Kendall1999JR, Guigui2021JR}, Poincare polydisc or hyperbolic space. Furthermore, \emph{geoopt} provides  optimizers like SGD or Adam and the sampling from a probability distribution on the manifold, whereas \emph{Geomstats} provides Frechet mean estimators, $K-$means, and principal component analysis.

\emph{Theseus} \cite{Pineda2022JR} provides \emph{differentiable} optimizers (Gauss-Newton, Levenberg-Marquardt) and solvers (dense and sparse versions of Cholesky and LU) as well as the manifolds $SO(3)$ and $SE(3)$ which are often used in 3D data processing, robotics and kinematics. The differentiability of the optimizers/solvers makes it possible to include them into a neural network layer or loss function.



\section*{Acknowledgment}
This work was supported by European Union´s Horizon 2020 research and innovation programme under grant number 951911 - AI4Media.

\bibliographystyle{abbrv}
\bibliography{sample}


\end{document}